\documentclass{osa-article}
\journal{preprint}
\articletype{Research Article}
\begin{document}

\title{High-sensitivity Frequency Comb Carrier-Envelope-Phase Metrology in Solid State High Harmonic Generation 
}
\author{Daniel M. B. Lesko,\authormark{1,2,3*} Kristina F. Chang,\authormark{1} and Scott A. Diddams\authormark{1,3,4,*}}
\address{\authormark{1}Time and Frequency Division, National Institute of Standards and Technology, Boulder, CO, USA\\
\authormark{2}Department of Chemistry, University of Colorado, Boulder, CO, USA\\
\authormark{3}Department of Physics, University of Colorado, Boulder, CO, USA \\
\authormark{4}Department of Electrical, Computer and Energy Engineering, University of Colorado, Boulder, CO, USA \\
}

\email{\authormark{*}Daniel.Lesko@colorado.edu, Scott.Diddams@nist.gov} 


\begin{abstract}
Non-perturbative and phase-sensitive light-matter interactions have led to the generation of attosecond pulses of light and the control electrical currents on the same timescale.
Traditionally, probing these effects via high harmonic generation has involved complicated lasers and apparatuses to generate the few-cycle and high peak power pulses needed to obtain and measure spectra that are sensitive to the phase of the light wave.
Instead, we show that nonlinear effects dependent on the carrier-envelope phase can be accessed in solid state crystals with simple low-energy frequency combs that we combine with high-sensitivity demodulation techniques to measure harmonic spectral modulations.
Central to this advance is the use of a scalable 100~MHz Erbium-fiber frequency comb at 1550~nm to produce 10~nJ, 20~fs pulses which are focused to the TW/cm$^2$ level. In a single pass through a 500~\textmu m ZnO crystal this yields harmonic spectra as short as 200~nm. 
With this system, we introduce a technique of carrier-envelope amplitude modulation spectroscopy (CAMS) and use it to characterize the phase-sensitive modulation of the ultraviolet harmonics with 85~dB signal-to-noise ratio.
We further verify the non-perturbative nature of the harmonic generation through polarization gating of the driving pulse to increase the effects of the carrier-envelope phase.
Our work demonstrates robust and ultra-sensitive methods for generating and characterizing harmonic generation at 100 MHz rates that should provide advantages in the study of attosecond nonlinear processes in solid state systems. Additionally, as a simple and low-noise frequency comb, this broadband source will be useful for precision dual-comb spectroscopy of a range of physical systems across the ultraviolet and visible spectral regions (200~-~650~nm). 
\end{abstract}

\section{Introduction}

Non-perturbative and phase-sensitive nonlinear optics opens new windows to physical phenomena in gases, liquids, and solids on the attosecond timescale\cite{Krausz2014}. 
These attosecond ($10^{-18}$s) processes in atomic\cite{Hentschel:2001,Sansone:2011,Chini:2014} and solid state semiconductor systems\cite{Garg2016,Ghimire_2014,Sederberg2020,Ghimire2019,You:17, Li2020, Higuchi:14} are most commonly probed by way of light emitted from high harmonic generation (HHG).
Conductors\cite{Boolakee2022,Bionta2021} and plasmonics\cite{Hommelhoff2006,Putnam2017,Piglosiewicz2014,PhysRevLett.109.244803,Heide2020,Kruger2011,Lemell2003} similarly exhibit these attosecond phenomena but are probed by phase sensitive current generation. 
Such ultrafast physical phenomena in both high and low bandgap materials can potentially be used for generating and controlling ultrafast transient electron currents for petahertz ($10^{18}$~Hz) electronics and exploring light matter interactions at attosecond timescales\cite{Krausz2014,Higuchi2017,Schultze2013,Schiffrin2013,Boolakee2022}.
Despite the large energy difference in optical fields required to drive these processes, the requirement of stable few-cycle optical pulses for exciting sub-cycle dynamics is common across this wide range of research topics.
Probing these dynamics require pulses with a well defined carrier envelope phase (CEP), i.e. a repeatable and controllable waveform, to provide a known electric potential on the timescale of the cycle of light.

Non-perturbative HHG processes in gases and semiconductors typically require high power lasers with \textmu J~to~m J energies and $\leq$10 kHz repetition rates. 
But similar phase dependent processes are present in solid state conductors which require significantly lower optical energies, $\leq$5 nJ, to probe and are therefore compatible with high repetition rate sources, $\ge$80 MHz.
To elucidate the phase sensitivity in the HHG process, the CEP is locked to a well defined value, and slowly scanned to measure the phase-dependent spectral shifts\cite{Hentschel:2001,Sansone:2011,Chini:2014,Sederberg2020,Ghimire2019,Ghimire_2014,You:17, Li2020, Higuchi:14}.
This technique relies on the high fluxes achieved with low repetition rate sources to produce measurable spectral shifts with a grating spectrometer.
On the other hand, when exploring these processes in conductors, one can take advantage of the high repetition rate and high frequency modulation and demodulation techniques to perform similar measurements\cite{Putnam2017,Boolakee2022}.
With a non-zero \textit{f}$_{\text{ceo}}$, the CEP cycles through 2$\pi$ at a well defined rate. 
By using a lock in at \textit{f}$_{\text{ceo}}$, and simple symmetry arguments\cite{Boolakee2022}, one can determine the phase dependent current generation as well as the contributions of the interband and intraband currents to the total signal.
The power of this technique relies in the high frequency modulation and demodulation allowing for high signal to noise measurements that can detect exceedingly small current modulations using patterned electrodes.
However, this approach relies on plasmonic enhancements in conductors and does not produce HHG light.
Addressing subcycle attosecond dynamics in HHG with high repetition rate sources requires a significant advancement in the methods used to detect these CEP dependent signals due to the low pulse energies of the generated HHG light.

Here we introduce high-sensitivity frequency comb techniques found in sub-cycle current generation in conductors to study phase-sensitive harmonic generation from solid dielectrics with low energy (nJ) pulses at 100 MHz.  
This advance is enabled by our technique of low-noise and scalable short pulse generation that overcomes conventionally limited powers from 100~MHz Er:fiber combs to produce 10~nJ 20~fs pulses at 1550~nm\cite{Lesko2021}.
We focus these pulses into 500~\textmu m ZnO (11-20) to 2~TW/cm$^{2}$, producing harmonics in a near continuum to wavelengths as short as 200~nm, without the need for high pressure hollow core fibers, pulse picking, or complicated vacuum apparatuses.
To measure CEP dependent spectral modulations we leverage the low noise properties of our Er:fiber comb to characterize extremely small amplitude modulation sidebands in the ultraviolet (UV) harmonics that arise from the nonzero $\textit{f}_{\text{ceo}}$.
This approach, which we call CAMS (carrier-envelope amplitude modulation
spectroscopy), provides 85~dB of signal-to-noise ratio (SNR) at a 1~Hz resolution bandwidth (RBW), allowing us to measure the effect of the CEP cycling in a 4 cycle pulse.
Analyzing multiple harmonics with CAMS, reveals the impact of the crystalline symmetry on the periodic  spectral modulations that arise from the pump CEP.
We further confirm the non-perturbative nature of our generated light by gating our pulse, effectively shortening it, and observing increased modulation and non-perturbative power scaling. 
The use of solid state target and a fiber laser system results in a simple, robust, and vacuum free apparatus to measure these strong field effects.
We anticipate systems like this will be useful not only for measuring field sensitive physics in solids and potentially gases, but also for broadband spectroscopy in a dual comb modality.

\section{Experimental Setup}
An outline of the experimental setup is shown in Figure 1a. The setup is based on a commercial, 100~MHz low noise polarization maintaining (PM) Er:fiber oscillator at 1550~nm (Menlo Systems). 
The $\textit{f}_{\text{ceo}}$ of the oscillator is stabilized by a conventional \textit{f}-2\textit{f} interferometer to a maser referenced signal at $\textit{f}_{\text{ceo}}=1$~MHz. 
This provides a well defined cycling of the CEP with the sampling defined by $(2\pi\textit{f}_{\text{ceo}})/(\textit{f}_{\text{rep}})$ $\sim 63$~mrad/pulse for the 100~MHz oscillator. 
The $\textit{f}_{\text{rep}}$ is not locked but maintains enough stability over a typical 10 second average to negligibly impact the rate of CEP cycling. 

The oscillator pulses are  amplified and spectrally broadened to support few cycle pulses, as described in Ref. \cite{Lesko2021}. Briefly, the oscillator output is stretched in a normal dispersion fiber and amplified to 20~nJ in a purpose-built erbium/ytterbium doped fiber amplifier. 
The pulses are then compressed with a grating compressor, spectrally broadened in PM normal dispersion highly nonlinear fiber (ND-HNLF), and compressed again using fused silica (FS) and third order dispersion mirrors.
The dispersion is optimized with the FS wedges for a near transform limited pulse at the back face of the generation crystal.
The result are 10~nJ, 20~fs, few-cycle pulses at the back face of crystal measured by second harmonic generation frequency resolved gating (SHG-FROG, Supplemental Figure 1).
The pulses are then tightly and achromatically focused by an off-axis parabolic mirror (OAP) to a 1/e$^2$ radius of 4.5 \textmu m in a 500~\textmu m thick a-plane single crystal ZnO (11-20). 
Our SHG-FROG measurements (supplemental Figure 1) indicate that linear compression in the ZnO dominates over nonlinear compression processes.
Spatially, the focus is placed at the back of the crystal to avoid re-absorption of light generated above the ZnO bandgap.
The peak power is estimated to be $\sim$0.675~MW corresponding to a peak intensity of $>$2~TW/cm$^2$.
While the peak power of the pulse is enough to consider self-focusing, the lack of observed self collapse and spectral broadening for the driving pulse would suggest this is not a dominant process for producing high intensities.
The dynamics between the possible self-focusing and nonperturbative harmonic generation would require further study.

The UV and visible (UV/Vis, 200~-~650~nm) light generated with the ZnO crystal is collected by an OAP with high reflectance in the UV (Acton \#1200, 120~-~600~nm), and sent to a purpose-built monochromator. 
Our monochromator is based on a 1800~g/mm grating blazed for 250~nm (Richardson Gratings) mounted on a rotation stage and a fast, UV-sensitive photo-multiplier tube (PMT, H6780-03 Hamamatsu). The estimated resolving power is $\lambda/\delta\lambda=125$  at 250~nm. The DC photocurrent produced by the PMT is amplified using a low noise current pre-amplifier (Femto) and measured simultaneously on an oscilloscope and spectrum analyzer. 
At each grating position, the dark photocurrent is measured and subtracted, while the $\textit{f}_{\text{ceo}}$ power is normalized to the $\textit{f}_{\text{rep}}$ power. 
The power spectral density is calibrated by measuring the the third harmonic yield with a notch filter and power meter. 
Due to detector saturation, the spectra of the $3^{rd}$ and $4^{th}$ harmonic are taken with a neutral density filter in line before being scaled to match the optical power measured and concatenated with the higher harmonics.
A UV Glan-Thompson polarizer is used in a rotation mount to measure the polarization of the generated light.

\begin{figure}[h!]
\centering\includegraphics[width=\textwidth]{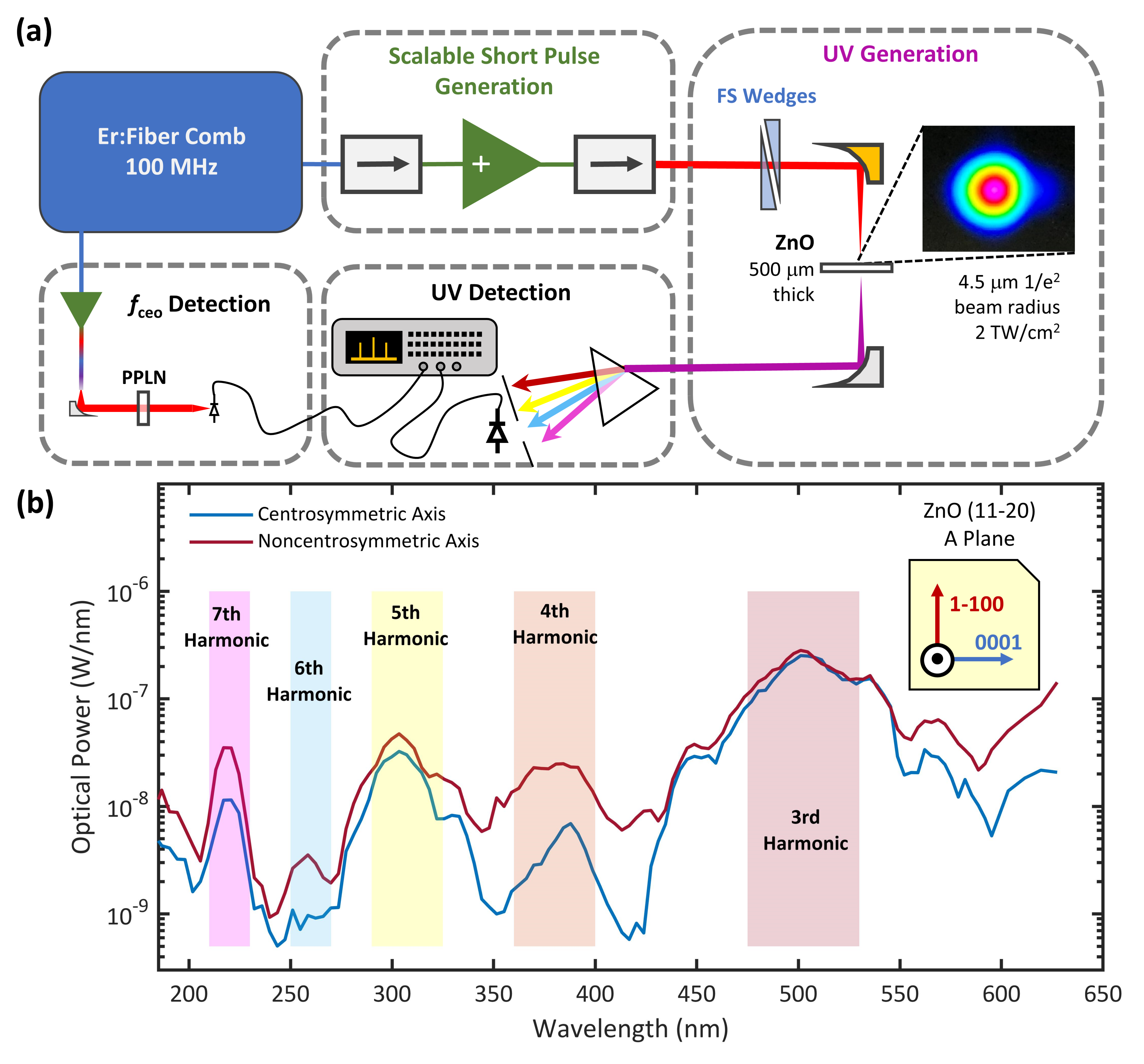}
\caption{Overview of solid state high harmonic generation (HHG) driven by a frequency comb: \textbf{(a)} Experimental setup utilizing short pulse generation at 1550 nm with a low-noise Er:fiber comb to produce high-power, 20~fs pulses. The pulses drive HHG in a 500~\textmu m thick, a-plane cut ZnO (11-20). Generated UV and visible light is detected by a monochromator and photomultiplier tube. \textbf{(b)} Spectra resulting from HHG in ZnO. HHG oriented along the centrosymmetric axis (0001, blue) yields predominately odd harmonics while the noncentrosymmetic axis (1-100, red) yields both even and odd harmonics. The peak at $\sim$385~nm appears to be consistent with photoluminescence on the centrosymmetric axis demonstrated in\cite{Hollinger20,Hollinger21}. 
}
\end{figure}

\section{Results}
Spectra of UV and visible light from 200-650 nm generated in the ZnO crystal are presented in Figure 1b. The cut of the ZnO crystal (a-plane, 11-20) enables the crystal to be oriented such that excitation primarily occurs along either the centrosymmetric axis (0001, blue) or the noncentrosymmetric axis (1-100, red).
Due to crystal symmetry, generation along the centrosymmetric axis predominantly yields odd harmonics of the fundamental 1550 nm driving laser, as well as photoluminesence at $\sim$385~nm\cite{Hollinger20,Hollinger21}. 
Some even harmonics are also observed along the centrosymmetric axis, but are significantly weaker relative to the odd harmonics. 
The observation of weak even harmonics may arise from harmonic generation at the crystal surface where the symmetry of the crystal is broken, from off-axis generation due the tight focus of the driving laser, or from co-propagating surface generated second harmonic.
In contrast, generation along the noncentrosymmetric axis yields both even and odd harmonics, resulting in more continuous spectral coverage compared to the centrosymmetric axis.
In both crystal orientations, generation up to the 7th harmonic ($\sim$221~nm) is observed.
Given the intensity of the driving pulse, harmonics beyond the 7th harmonic are expected\cite{Wang2017,Liu:17}, but lie beyond the wavelength range of the detector.
In both crystal orientations, the harmonics are observed to be primarily co-polarized from the driving laser field. 
The observation of co-polarized harmonics is consistent with previous HHG experiments in ZnO\cite{Ghimire2011,Jiang_2019}.
With the 10~nJ, 100~MHz, driving laser, >10$^{10}$ photons/second/nm are produced at 221~nm, the highest observed harmonic (see Supplemental Figure 2).

With a short and tightly focused pulse, we begin to observe CEP dependent spectra, similar to systems with much higher pulse energies and lower repetition rates.
To detect and measure this effect, we spectrally filter the generated light using a monochromator and detect the RF spectrum with a high bandwidth UV sensitive PMT (Figure 1a).
By having an $\textit{f}_{\text{ceo}} \neq 0$, the CEP cycles at a well defined rate and changes the harmonic generation process from pulse to pulse (Figure 2a). 
The change in the CEP is imprinted onto the spectrum (Figure 2b(i)) and measured in the time domain as modulations on the repetition rate of the comb (Figure 2b(ii)).
The Fourier transformation of the monochromator signal reveals the amplitude modulation depth ($\beta$) and frequency, which is proportional to the rate at which the CEP cycles (i.e. $\textit{f}_{\text{ceo}}$). 
This method of Carrier-envelope Amplitude Modulation Spectroscopy (CAMS) allows for narrow resolution bandwidth (1~Hz) to measure modulations as small as -85~dBc, relative the $f_{rep}$ tone, as shown Figure 2c.
Notably, we only observe $\textit{f}_{\text{ceo}}$ and 2$\textit{f}_{\text{ceo}}$ in the RF spectra, corresponding to the periodic dependence (2$\pi$ and $\pi$ respectively) of the UV light on the CEP.
The effect of the CEP periodicity is seen in Figure 2d. 
When the noncentrosymmetric axis is used the $\textit{f}_{\text{ceo}}$ tone is significantly increased.
This is due to the lack of degeneracy between a CEP of 0 and $\pi$ (i.e. a cosine and -cosine pulse) in the noncentrosymmetric material.
UV generation on the centrosymmetric axis results in a suppression of the $\textit{f}_{\text{ceo}}$ tone.
The $\textit{f}_{\text{ceo}}$ does not disappear due to the small amount of surface generated second harmonic\cite{Li2020} and tight focusing.
Both of these effects slightly break the degeneracy of the two CEP values (0 and $\pi$) resulting in $\textit{f}_{\text{ceo}}$ being present on both axes.

Narrow-band RF detection allows the possibility for two additional measurements that can be taken. 
With a lock-in detection scheme, the phase difference between the amplitude modulations at $\textit{f}_{\text{ceo}}$ and 2$\textit{f}_{\text{ceo}}$ can be measured as a function of wavelength (Supplemental Figure 3). 
This measurement could potentially be used to give information on the chirp of the UV pulses, similar to a traditional CEP scan using a grating spectrometer\cite{You:17}.
In future experiments, without spectral filtering by a grating monochrometer, one could observe the center of mass shifts (timing jitter) of the UV pulses relative to the driving laser envelope.
The center of mass shift could yield information about the phase delay of the generated UV light with respect to the driving laser's CEP.
Here, the analysis at higher harmonics of the $\textit{f}_{\text{rep}}$ could be beneficial, where the timing jitter has a stronger impact on the signal than the amplitude modulation.

\begin{figure}[h!]
\centering\includegraphics[width=\textwidth]{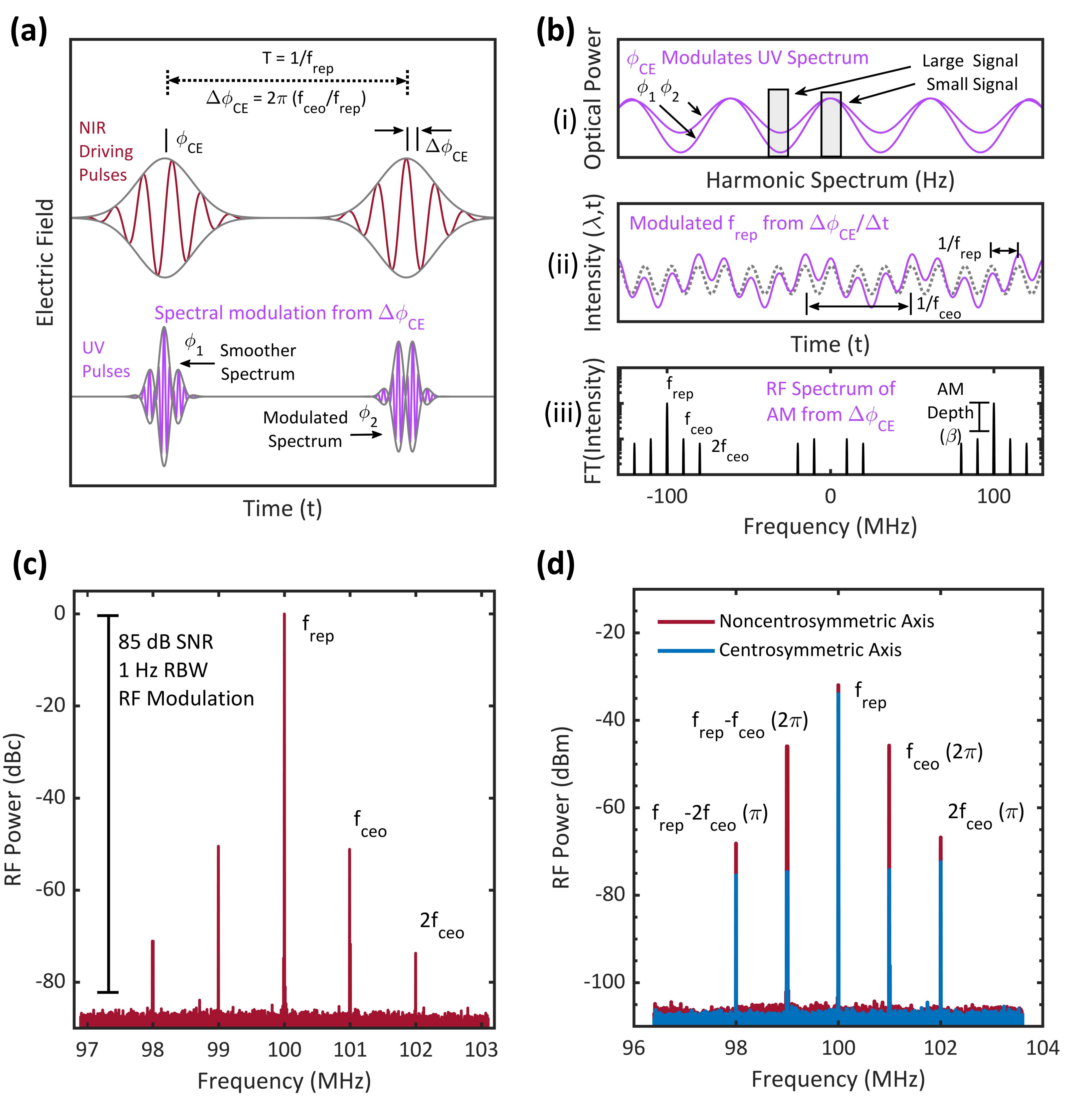}
\caption{Amplitude modulation from carrier envelop phase (CEP) dependent harmonic generation measured by Carrier-envelope Amplitude Modulation Spectroscopy (CAMS).
\textbf{(a)} Schematic describing the timing of the driving laser pulses with CEP cycling, as well as the resulting CEP dependent UV generation. 
\textbf{(b)} Schematic describing the detection method of the CEP dependent spectrum,
(i) The CEP's dependence can be seen across the spectra, with certain regions yielding larger signals. 
(ii) Isolating one wavelength region, this CEP dependent spectral intensity can be seen as modulation on the \textit{f}$_{\text{rep}}$ tone.
(iii) A Fourier transform reveals the modulation depth of the CEP dependent intensity ($\beta$) as well as the frequency as well as the CEP cycling frequency (\textit{f}$_{\text{ceo}}$).
\textbf{(c)} At the center of harmonics, (data from the third harmonic, $\sim$500~nm) we achieve $>$85 dB sensitivity to CEP effects allowing for measurements of small modulations. We observe \textit{f}$_{\text{ceo}}$ and 2\textit{f}$_{\text{ceo}}$ tones in the spectrum. 
\textbf{(d)} RF spectra at 439~nm showing the effect of symmetry breaking on the relative amplitudes of \textit{f}$_{\text{ceo}}$ and 2\textit{f}$_{\text{ceo}}$. The presence of \textit{f}$_{\text{ceo}}$ and 2\textit{f}$_{\text{ceo}}$ corresponds to a 2$\pi$ and $\pi$ periodicity of the signal on the CEP, respectively.}
\end{figure}

With CAMS, we measure the modulation depth $\beta$ across the spectrum (from 200 nm~to~700 nm) for both the centrosymmetric (Figure 3a) and noncentrosymmetric (Figure 3b) axes.
As noted above, the increased symmetry breaking on the noncentrosymmetric axis gives much larger $\beta$ 
across the spectrum corresponding to increased modulation with a 2$\pi$ periodicity.
On both crystallographic axes, the lower order harmonic's modulation depth follows a trend of being more prominent at the wings of each harmonic and sharply diminished at the center. 
This is because at a $\sim$3.9 cycle pulse length, the positions of the harmonic centers are: i) not shifting dramatically with change of CEP, and ii) being averaged over the entire 2$\pi$ of the CEP by our integration time (set by the $\textit{f}_{\text{ceo}}$ and resolution bandwidth).
One clue as to the probable nonperturbative character of the generated spectra is the flat modulation depth between the 5th and 7th harmonic on the centrosymmetric, where only $\textit{f}_{\text{ceo}}$ is present, despite there being no distinguishable 6th harmonic. 
One would expect to see 2$\textit{f}_{\text{ceo}}$ between these harmonics from heterodyne gain (a 5\textit{f} - 7\textit{f} interference) if this was a perturbative process. 
Furthermore, we do not observe any $\textit{f}_{\text{ceo}}$ tone anywhere on the 1550 nm fundamental, which would originate from a cascaded $\chi^{(2)}$ (perturbative) process.

\begin{figure}[h!]
\centering\includegraphics[width=\textwidth]{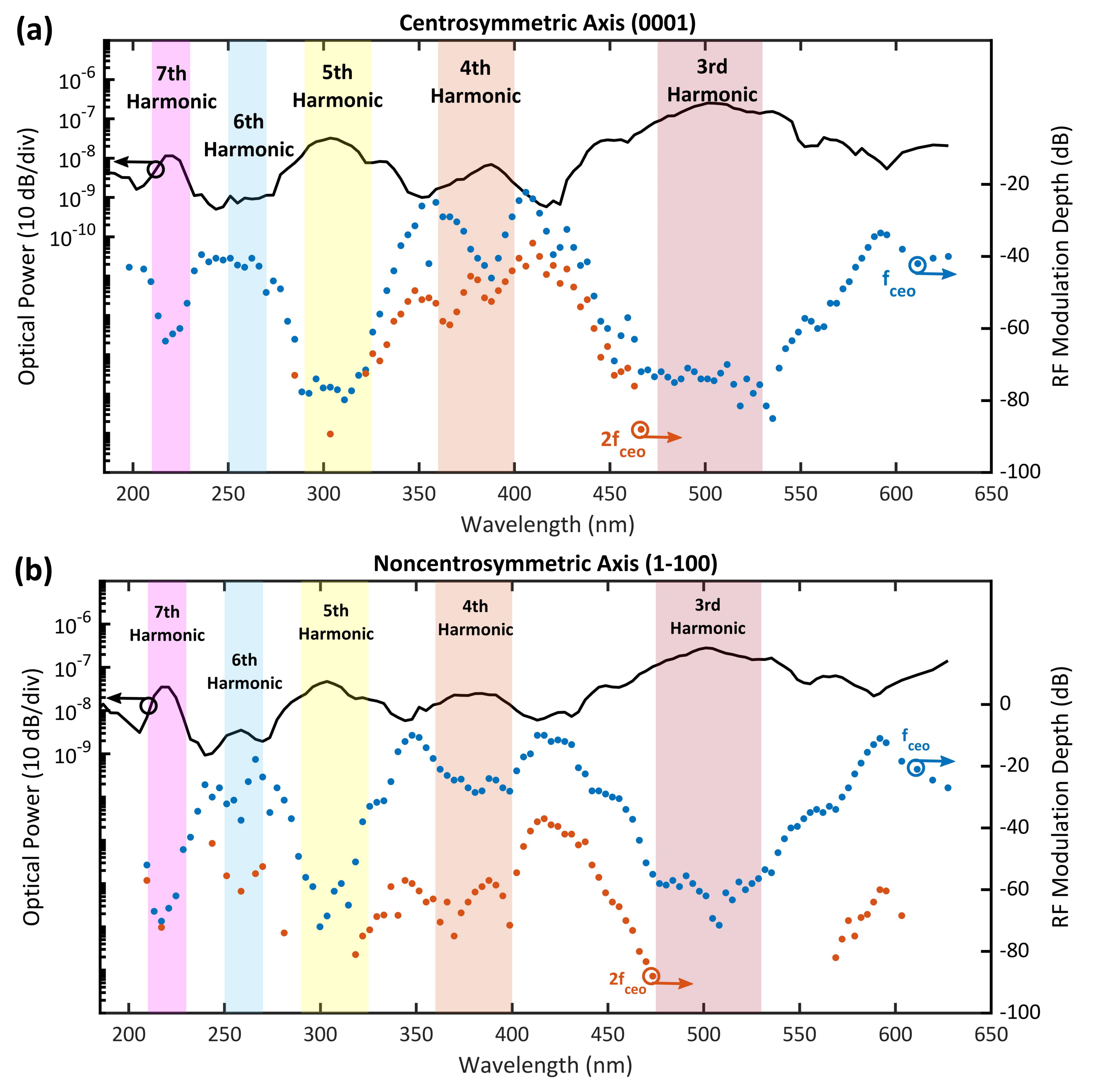}
\caption{CAMS spectra of measured RF power modulation (shown as n\textit{f}$_{\text{ceo}}$/\textit{f}$_{\text{rep}}$) across the UV/Vis spectrum on the centrosymmetric \textbf{(a)} and noncentrosymmetric \textbf{(b)} axes.
}
\end{figure}

To further investigate these arguments of perturbative and non-perturbative processes generating the UV/Vis light, we use polarization assisted amplitude gating (PASSAGE)\cite{Timmers:16} (Figure 4a) to slightly reduce the number of cycles that contribute to the nonlinear process.
To implement PASSAGE, two achromatic $\lambda/2$ waveplates impart a $\lambda$ shear, while an achromatic $\lambda/4$ waveplate imparts ellipticity.
We compensate the additional dispersion of the waveplates by adding/removing UVFS glass.
The time-varying polarization effectively reduces the number of cycles contributing to the generation of UV light.   
The impact of such polarization gating has been observed with gases, which exhibit a strong dependence on polarization\cite{Timmers:16}.
However, solids such as MgO and ZnO exhibit anisotropy, and similar polarization dependent effects on harmonic generation are also expected\cite{Ghimire2011,Jiang_2019,Hollinger21}.
When we measure the modulation depth (\textit{f}$_{\text{ceo}}$/\textit{f}$_{\text{rep}}$) as a function of the $\lambda$/4 plate angle, we see an increase of 23.3~dB at 45~degrees (Figure 4b).
Furthermore, the shape of the curve, and singular peak of the ellipticity on the modulation depth (Figure 4b) suggests that a time-dependent elliptical profile on the driving pulse is largely maintained, despite the birefringence of the a-cut ZnO crystal\cite{Hollinger21,Jiang_2019}.
Since the efficiency of HHG in ZnO, and most solids in general, possesses weaker sensitivity to the polarization\cite{Hollinger21,Jiang_2019} of the driving pulse than gas phase HHG, the shape of the modulation depth as a function of ellipticity is present but not sharp.
Nonetheless, our measurements show that PASSAGE, and more generally polarization gating, can still be applied to solid state HHG in ZnO, yielding an increased sensitivity to the driving laser CEP. 
The overall yield of UV light is decreased by approximately a factor of 8, which is in agreement to scaling seen in the gas\cite{Timmers:16}, and consistent with a non-perturbative picture.
In a perturbative picture, the 5th harmonic light would be reduced by a factor of $>$10000 based on a simple intensity$^5$ scaling from the reduction of the driving pulse, far below our detection range.

To further show that the generated UV light has non-perturbative character, we measure the change in modulation depth ($\Delta\beta$) from PASSAGE across part of the UV spectrum (Figure 4c).
If our generation of UV light was governed by perturbative nonlinear optics, we would expect the harmonics to increase in width due to the effectively shorter (sub-cycle) driving pulse.
The $\Delta\beta$, normalized heterodyne gain from n\textit{f} - (n-1)\textit{f} interference, from this spectral widening would correspond to a increase in signal closer to the center of each harmonic.
This would arise from the increase in spectral overlap closer to the center of the harmonics.
However, this trend in $\Delta\beta$ is not observed between the 5th and 7th harmonics (Figure 4c). 
Deviation in the observed $\Delta\beta$ trend from that expected for a purely perturbative harmonic generation mechanism provides further evidence for a non-perturbative generation mechanism.\par

\begin{figure}[h!]
\centering\includegraphics[width=\textwidth]{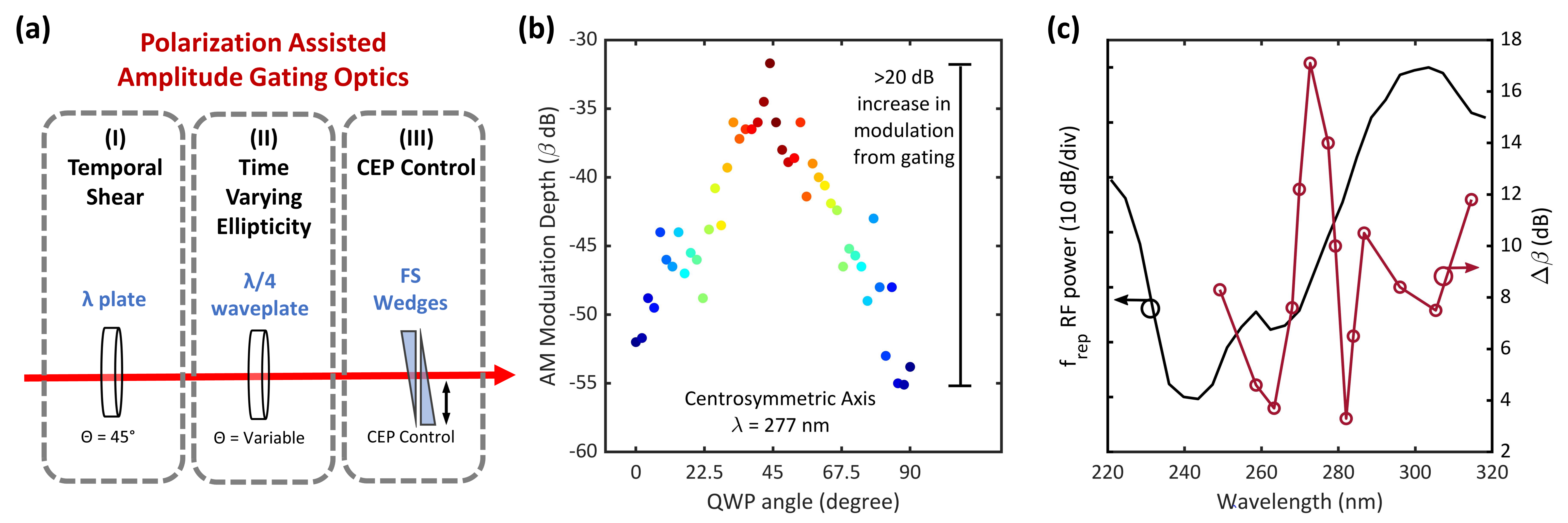}
\caption{Increasing the carrier envelope phase effects in ZnO. \textbf{(a)} We utilize polarization assisted amplitude gating (PASSAGE) to reduce the number of cycles contributing to UV generation. By introducing a (I) temporal shear by a $\lambda$ plate we are able to impart a time dependent ellipticity (II) by a $\lambda$/4 plate. This reduces the effective driving pulse to a $<$1 cycle pulse. Finally we can optimize the CEP and chirp by FS wedges (III). \textbf{(b)} Modulation depth ($\beta$) as a function of the $\lambda$/4 angle, showing a 23.5~dB increase in the UV modulation depth when fewer cycles are used in the generation process. \textbf{(c)} Measured increase in modulation ($\Delta\beta$) from using PASSAGE on the centrosymmetric axis between the 7th and 5th harmonics.The shape of the increase in modulation does not match a perturbative picture of widening harmonics.}
\end{figure}

The CEP-sensitive spectral modulations detected by CAMS are measured with $>$85 dB (at 1 Hz RBW) of dynamic range. 
The sensitivity afforded by CAMS exceeds that of a traditional UV spectrometer based on commercially available cameras\cite{fellers_davidson} by several orders of magnitude. 
Leveraging this high sensitivity, one can envision extending the technique to study the CEP dependent process in semiconductors, where the CEP dependent signal would be drastically smaller for the intraband currents than the interband currents\cite{You:17}.
Furthermore, one could probe materials that are thought to not have CEP sensitivity in their interband or intraband currents\cite{You22017}.
These intraband and interband current phenomena have been studied in conductor systems\cite{Boolakee2022} with similar lock-in techniques, but they have not been studied with such sensitivity in semiconductor solids and gases.

\section{Conclusions}
In summary, we present non-perturbative single pass solid state HHG based on a robust, low noise, and compact 100 MHz Er:fiber frequency comb. 
We utilize high frequency modulation/demodulation techniques to measure the spectral modulation from the CEP cycling with 85~dB SNR.
This allows us to measure the spectral modulations from  CEP sensitive HHG UV pulse shearing with a 4 cycle pulse.
This simple, robust, high intensity and high repetition rate source will be useful for investigating field sensitive physics in semiconductor solids and gases that benefit from detection of weak signals and the intrinsic fast averaging at the 100 MHz rate.
Furthermore, the broadband UV/Vis spectrum that is generated with the noncentrosymmetric axis of ZnO will match broadband atmospheric UV absorbers such as NO$_2$ and SO$_2$ for dual comb spectroscopy.
With the amount of light generated at the 5th harmonic, we estimate it will be possible to measure spectra across 100 THz of optical bandwidth with 10 GHz resolution for averaging times <1 hour. 
Work towards such experiments, including the construction of a second frequency comb system, is ongoing.

\begin{backmatter}
\bmsection{Funding}
This research was supported by the Defense Advanced Research Projects Agency SCOUT Program, the Air Force Office of Scientific Research (FA9550-16-1-0016) and NIST.

\bmsection{Acknowledgments}
Mention of specific products or trade names is for technical and scientific information and does not constitute an endorsement by NIST.
D.M.B.L. acknowledges award 70NANB18H006 from NIST. 
K.F.C. acknowledges support from the National Research Council.
The authors thank J. Ye and C. Zhang for loaning VUV optics, as well as T. Allison, J. Biegert, M. Chini, S. Ghimire and F. Quinlan, for valuable comments and discussion.

\bmsection{Disclosures} The authors declare no conflicts of interest.
\bmsection{Data availability} Data underlying the results presented in this paper are not publicly available at this time but may be obtained from the authors upon reasonable request.


\end{backmatter}

\bibliography{sample}

\section{Supplemental Information}

\subsection{Spectral Phase Compensation in ZnO}
Our scalable short pulse generation\cite{Lesko2021} produces pulses with bandwidths able to support few-cycle pulses. 
The pulses are compressed out of the polarization maintaining normal dispersion highly nonlinear fiber (PM-ND-HNLF, OFS, -2.9 ps/nm/km, 0.014ps/nm$^2$/km) with fused silica (for second order phase compensation) and third order dispersion compensation mirrors (Ultrafast Innovations, -750 fs$^3$). 
Since the harmonics are generated near the back face of the ZnO, fused silica is removed from the beam path to yield a chirped pulse at the entrance of the crystal measured by second harmonic generation frequency resolved optical gating (SHG-FROG, Figure 5a).
As the pulse propagates in the ZnO the spectral phase is compensated for by the material dispersion, producing a 20 fs pulse at the output of the crystal (Figure 5b).
This spectral phase (amount of glass added) is optimized by maximizing the 7th harmonic observed.
The reconstructed pulse temporal profile and spectral phases of the pulse before and after are shown in Figures 5c and 5d.
While the vacuum intensity is $~$2~TW/cm$^2$, self compression is not dominant shown in the lack of spectral broadening in the retrieved pulses' spectra (Figure 5e).
Independent optical spectra from a commercial optical spectrum analyzer do not show substantial spectral broadening.

\begin{figure}[h!]
\centering\includegraphics[width=0.75\textwidth]{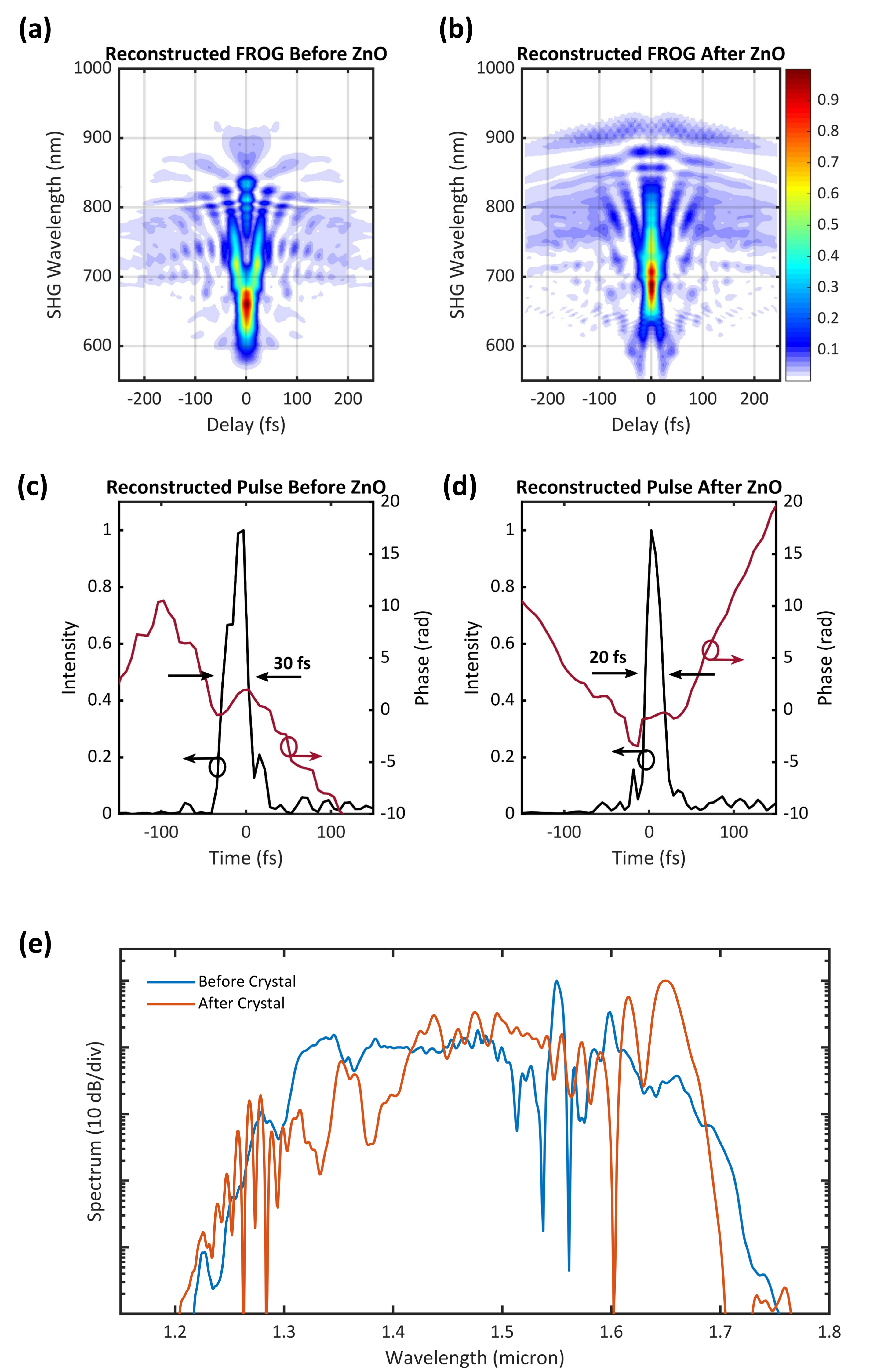}
\caption{Few-cycle pulse compression in ZnO measured by SHG-FROG. a) Pulse before ZnO is chirped to 30~fs. b) Pulse after ZnO has compressed to 20~fs from the material dispersion present. 
}
\end{figure}

\subsection{HHG Source Brilliance}
To provide a better comparison of our 100 MHz Er:fiber comb to other UV sources, we calculate the photons/s/nm (Figure 6) from the power spectral density (main paper Figure 1).
A factor of $10^{-8}$ s/pulse gives the brilliance in photons/pulse/nm.

\begin{figure}[h!]
\centering\includegraphics[width=0.75\textwidth]{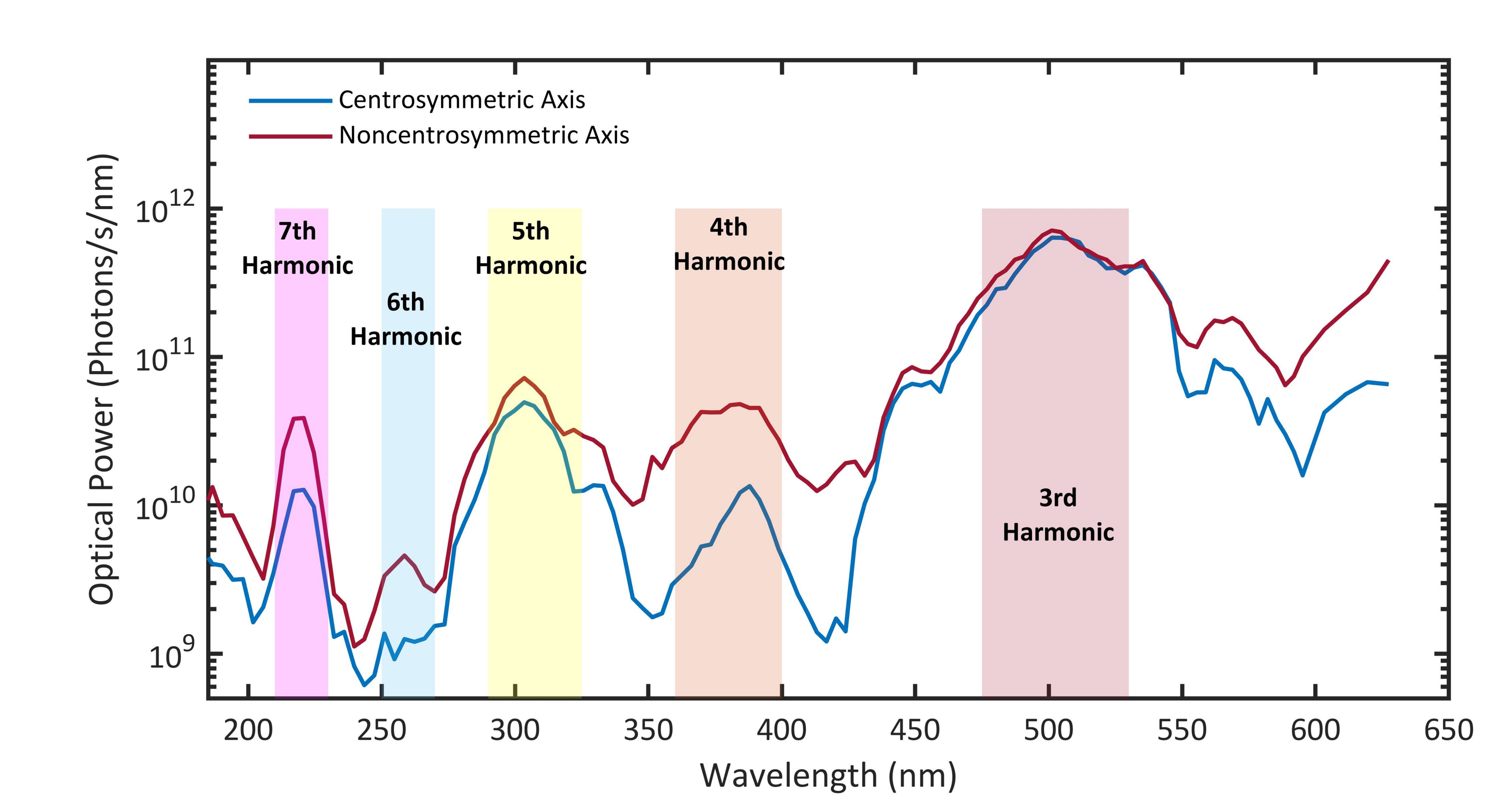}
\caption{
Measured brilliance of the light produced from high harmonic generation in 500 \textmu m 11-20 ZnO. The centrosymmetric and noncentrosymmetric axes corresponds to the 0001 and the 1-100 crystallographic axes respectively. 
}
\end{figure}

\subsection{Periodicity Phase Measurement}
Due to the nonperturbative mechanism of the harmonic generation, we observe modulation of the spectrum corresponding to the carrier envelope phase (CEP).
Utilizing the high frequency modulation/demodulation technique outlined in the paper, we can observe harmonics of the CEP cycling frequency (\textit{f}$_{\text{ceo}}$, Figure 7a).
These harmonics correspond to the periodicity of the spectral modulation as a function of the CEP. 
Utilizing a lock-in detector, we can measure the relative phases of the two signals (Figure 7b and c).
We can measure a phase difference of 0.92~rad between the different axes of the crystal. 
By taking this measurement across the spectrum, one should be able to reconstruct the atto-chirp of the pulse similar to a traditional CEP scan done in HHG.

\begin{figure}[h!]
\centering\includegraphics[width=0.75\textwidth]{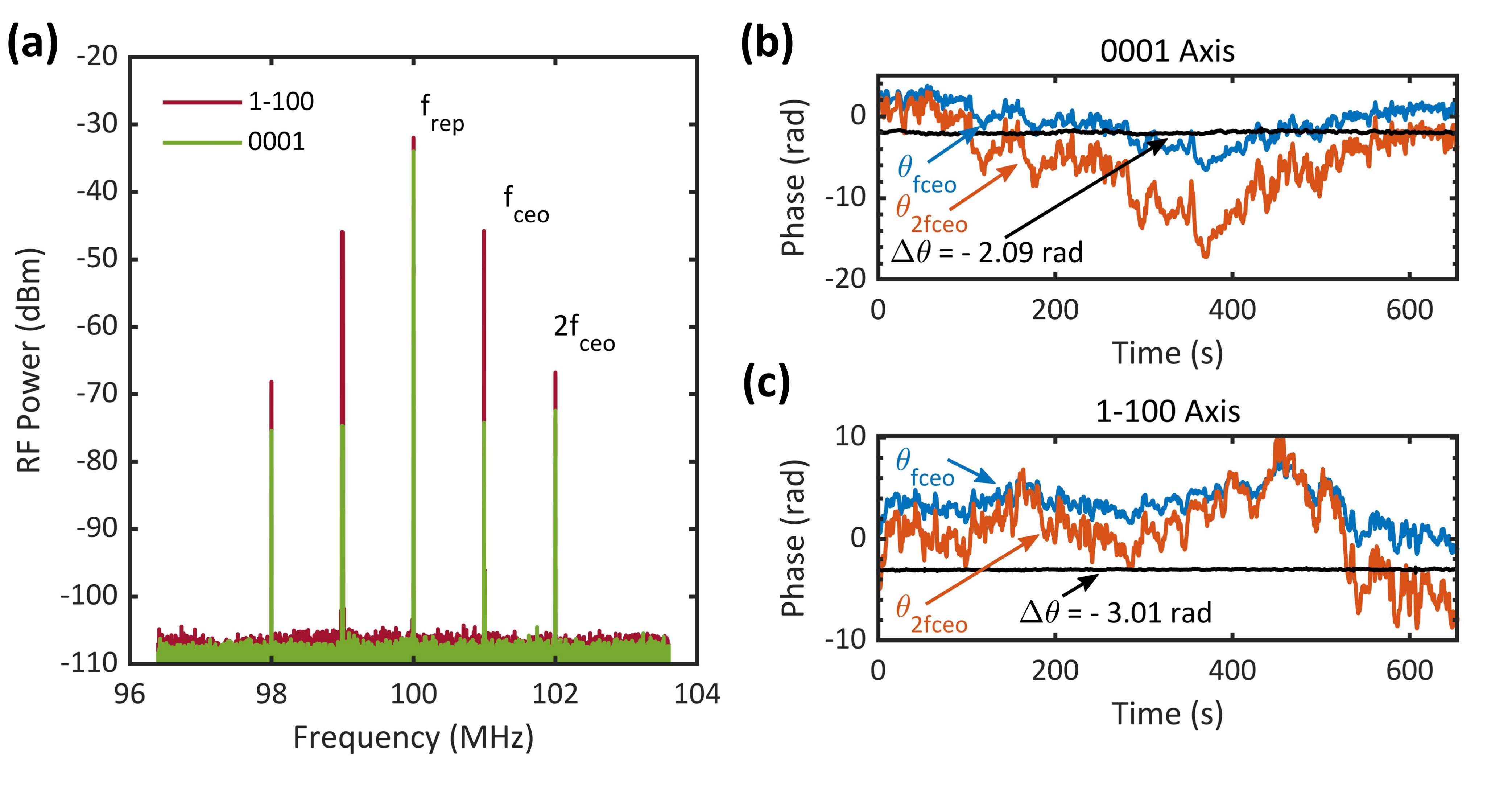}
\caption{Lock-in detection measurement of CEP dependent spectrum. (\textbf{a}) RF spectrum of the two \textit{f}$_{\text{ceo}}$ tones for the 1-100 and 0001 axes of ZnO. (\textbf{b}-\textbf{c}) Correlated phases on the 0001 and 1-100 axes result in a constant phase difference ($0.5*\theta_{\textit{2fceo}}-\theta_{\textit{fceo}}$).
}
\end{figure}

\end{document}